\documentclass[manuscript]{aastex}

\shorttitle{Solar cycle variation of the diurnal anisotropy}
\shortauthors{K.~Munakata et al.}

\begin{document}


\title{Solar cycle dependence of the diurnal anisotropy of 0.6 TeV cosmic ray intensity observed with the Matsushiro underground muon detector}


\author{K.~Munakata\altaffilmark{1}, Y.~Mizoguchi\altaffilmark{1}, C.~Kato\altaffilmark{1}, S.~Yasue\altaffilmark{1}, S.~Mori\altaffilmark{1}, M.~Takita\altaffilmark{2}, and  J.~Kota\altaffilmark{3}}
\email{kmuna00@shinshu-u.ac.jp}


\altaffiltext{1}{Department of Physics, Shinshu University, Matsumoto 390-8621, Japan.}
\altaffiltext{2}{Institute for Cosmic Ray Research, University of Tokyo, Kashiwa 277-8582, Japan.}
\altaffiltext{3}{Lunar and Planetary Laboratory, University of Arizona, Tucson, AZ 87721, USA.}


\begin{abstract}
We analyze the temporal variation of the diurnal anisotropy of sub-TeV cosmic ray intensity observed with the Matsushiro (Japan) underground muon detector over two full solar activity cycles in 1985-2008. We find an anisotropy component in the solar diurnal anisotropy superimposed on the Compton-Getting anisotropy due to the earth's orbital motion around the sun. The phase of this additional anisotropy is almost constant at $\sim$15:00 local solar time corresponding to the direction perpendicular to the average interplanetary magnetic field at the earth's orbit, while the amplitude varies between a maximum (0.043$\pm$0.002 $\%$) and minimum ($\sim$0.008$\pm$0.002 $\%$) in a clear correlation with the solar activity. We find a significant time lag between the temporal variations of the amplitude and the sunspot number and obtain the best correlation coefficient of +0.74 with the sunspot number delayed for 26 months. We suggest that this anisotropy might be interpreted in terms of the energy change due to the solar-wind induced electric field expected for GCRs crossing the wavy neutral sheet. The average amplitude of the sidereal diurnal variation over the entire period is 0.034$\pm$0.003 $\%$, which is roughly one third of the amplitude reported from AS and deep-underground muon experiments monitoring multi-TeV GCR intensity suggesting a significant attenuation of the anisotropy due to the solar modulation. We find, on the other hand, only a weak positive correlation between the sidereal diurnal anisotropy and the solar activity cycle, in which the amplitude in the ``active'' solar activity epoch is about twice the amplitude in the ``quiet'' solar activity epoch. This implies that only one fourth of the total attenuation varies in correlation with the solar activity cycle and/or the solar magnetic cycle. We finally examine the temporal variation of the ``single-band valley depth'' (SBVD) quoted by the Milagro experiment and, by contrast with recent Milagro's report, we find no steady increase in the Matsushiro observations in a 7-year period between 2000 and 2007. We suggest, therefore, that the steady increase of the SBVD reported by the Milagro experiment is not caused by the decreasing solar modulation in the declining phase of the 23rd solar activity cycle. 
\end{abstract}





\section{Introduction}
The sidereal anisotropy of high-energy galactic cosmic ray (GCR) intensity provides us with unique information about the magnetic structure of the heliosphere and/or the local interstellar space surrounding the heliosphere, through which GCRs propagate to the earth. Examining the temporal variation of the anisotropy is also of interest, since such a variation, if any, may reflect the change in the global magnetic structure in response to the solar  activity- and/or magnetic-cycle variations. While recent observations with air shower arrays and underground muon detectors have reported a consistent global structure of the anisotropy, conclusive reports on its temporal variation are still limited \citep[e.g., see][]{Nagashima89,Nagashima04,Amenomori05,Guillian07,Abdo09,Amenomori09}. The galactic anisotropy is measured as a sidereal daily variation (SDV) of cosmic-ray intensity recorded in a fixed directional channel of the detector on the spinning earth. The temporal variation of the SDV in connection with the solar magnetic cycle was investigated from a continuous observation of the SDV of $\sim$10 TeV cosmic-ray intensity over 12 years between 1973 and 1987 with an air-shower (AS) detector at Mt. Norikura in Japan. While the SDV in each year was statistically significant, its temporal variation was quite small with a statistical significance of only two sigmas \citep{Nagashima89}. Comparing data divided into two 5-year periods, 1997-2001 including the solar maximum and 2001-2005 approaching the solar minimum of the 23rd solar activity cycle, the Tibet III air shower experiment also concluded that the variation of 3 TeV cosmic ray anisotropy during the solar activity cycle was insignificant \citep{Amenomori06}. In the present paper, we analyze the long-term variation of the SDV by analyzing 0.6 TeV cosmic ray data observed with a Japanese underground muon detector over 24 years between 1985 and 2008. This observation period is long enough to examine the variation of both the 11-year solar activity- and 22-year solar magnetic-cycles.

Recently, the Milagro experiment utilizing a large water Cherenkov detector surrounded by an air shower array, on the other hand, has reported on a steady increase in the magnitude of the anisotropy of 6 TeV cosmic ray intensity over 7 years between 2000 and 2007 \citep{Abdo09}. This period corresponds to the declining phase of the 23rd solar activity cycle when the yearly mean sunspot number decreased from $\sim$100 to $\sim$10 toward a long-lasting activity minimum epoch. The increase reported by the Milagro experiment was so large that the anisotropy amplitude more than doubled between 2000 and 2007. If this increase did indeed, result from the decreasing solar activity, it is reasonable to expect that the effect should be observable also in the sub-TeV region, because the physical processes responsible for the solar modulation, such as adiabatic deceleration, pitch angle scattering by the magnetic irregularities and the solar wind convection effects in the heliosphere, are all expected to be more effective for lower energy cosmic rays. It has actually been established that the amplitude of the sidereal anisotropy decreases with decreasing energy between $\sim$0.1 and $\sim$1 TeV due to a larger attenuation by the solar modulation effect, while the amplitude is more constant for multi-TeV cosmic rays with energies over $\sim$1 TeV, implying insignificant influence of the solar modulation on TeV GCRs \citep{Munakata97,Nagashima98,Amenomori05,Amenomori09a}. In the present paper, we also examine whether the effect reported by Milagro is observed in the sub-TeV cosmic ray anisotropy.

The outline of this paper is as follows. In \S 2, we briefly describe the data and analysis method and present analysis results. We first show in \S 2.1 how the diurnal anisotropy in solar time changes in a significant correlation with the solar activity, indicating that solar modulation effects act on sub-TeV GCRs. We then show in \S 2.2 the long-term variation in the sidereal anisotropy. Our results are discussed in \S 3.

\section{Data and analysis}
The Matsushiro underground muon detector (36.53$\degr$N, 138.01$\degr$E, $\sim$300 km north-west of Tokyo) has been in operation since 1984. With its vertical overburden of 220 meters water equivalent, its muon energy threshold is 100 GeV. The multi-directional muon detector (sometimes also referred to as Zohzan according to the name of the hill above the observatory) consists of two horizontal layers of plastic scintillators, vertically separated by 1.5 m. Each layer comprises a 5$\times$5 square array of 1 m$^2$ unit detectors, each with a 1 m$\times$1 m$\times$0.1 m plastic scintillator viewed by two photomultiplier tubes (PMTs) of 12.7 cm diameter. In order to reduce the influence of the dark current noises, each detector is designed to output signals only when two PMTs simultaneously yield pulses each shaped with a width of 300 nanosecond. By counting pulses of the twofold coincidences between a pair of detectors on the upper and lower layers, we record the rate of muons from the corresponding incident direction. The multidirectional muon telescope comprises various combinations between the upper and lower detectors. In this paper, we analyze hourly muon rates in the eastern (E) and western (W) viewing channels comprised of 17 directional channels available from the detector. The average count rates of E and W channels are 7.14 Hz and 9.89 Hz, respectively, while the median primary cosmic ray energies recorded in these two composite channels are respectively calculated to be 0.660 TeV and 0.549 TeV on the basis of the response function of the atmospheric muons to the primary cosmic ray \citep{Murakami79}. The median energy of primary cosmic rays recorded in the vertical channels is also calculated to be 0.596 TeV.

In order to properly eliminate the spurious variations due to the atmospheric effect, we adopt the following ``East-West'' method which is similar to the ``Forward-Backward" method applied to Milagro data \citep{Abdo09}. Let $I(t)$ be the fractional deviation of the intensity in a vertical directional channel at the time $t$ in hours, defined as,
\begin{equation}
I(t)=[N(t)-\bar{N}]/\bar{N},
\end{equation}
where $N(t)$ is the hourly muon count and $\bar{N}$ is the 24-hour central moving average of $N(t)$ at $t$. Also let $R(t)$ be the percent deviation due to the anisotropy expected in the vertical directional channel. By taking the difference between $I(t)$s in the East- and West-viewing channels, $I^{E}(t)$ and $I^{W}(t)$, we can deduce the ``differential'' variation of $R(t)$ as,
\begin{equation}
D(t)=dR(t)/dt=[I^{E}(t)-I^{W}(t)]/\Delta t,
\end{equation}
where $\Delta t$ is the hour angle separation between the mean East- and West-incident directions averaged over the East- and West-incident muons and $I^{E}(t)$ and $I^{W}(t)$ are respectively composed of five directional channels \citep{Munakata06}. By utilizing the response function of the atmospheric muons to primary cosmic rays together with the particle trajectory code in the geomagnetic field \citep{Lin95}, we calculate $\Delta t$ to be 5.3 hours. Note that the ``differential'' variation $D(t)$ in eq.~(2) is free from the spurious variation due to the atmospheric effects which is expected to be common for $I^{E}(t)$ and $I^{W}(t)$. We consider an expansion of $R(t)$ into a Fourier series expressed with the amplitude $a^{R}_{n}$ and the phase $\phi^{R}_{n}$, as
\begin{equation}
R(t)=\sum_{n=1}^{3} a^{R}_{n} \cos n \omega (t - \phi^{R}_{n}),
\end{equation}
where $\omega = \pi /12$. By introducing this $R(t)$ into eq.~(2), we obtain $D(t)$, as
\begin{equation}
D(t)=\sum_{n=1}^{3} a^{D}_{n} \cos n \omega (t - \phi^{D}_{n}),
\end{equation}
where the coefficients $a^{D}_{n}$ and $\phi^{D}_{n}$ are respectively related to $a^{R}_{n}$ and $\phi^{R}_{n}$, as
\begin{equation}
a^{R}_{n}=a^{D}_{n}/(n \omega),
\end{equation}
\begin{equation}
\phi^{R}_{n}=\phi^{D}_{n}+6.0.
\end{equation}
We deduce $a^{D}_{n}$ and $\phi^{D}_{n}$ every year by best-fitting eq. (4) to the observed yearly mean $D(t)$, as
\begin{eqnarray}
a^{D}_{n}=\sqrt{(A^{D}_{n})^2+(B^{D}_{n})^2}, \\
\phi^{D}_{n}={\tan}^{-1} (B^{D}_{n}/A^{D}_{n}),
\end{eqnarray}
where
\begin{eqnarray}
A^{D}_{n}=(\omega/\pi) \sum^{24}_{i=1} D(t_{i}) \cos \omega t_{i}, \\
B^{D}_{n}=(\omega/\pi) \sum^{24}_{i=1} D(t_{i}) \sin \omega t_{i}.
\end{eqnarray}
Errors of $a^{D}_{n}$ and $\phi^{D}_{n}$ are propagated from the count rate error $\delta_{i}$ of $D(t_{i})$, as
\begin{eqnarray}
\delta a^{D}_{n} = \sqrt{(A^{D}_{n} \delta A^{D}_{n})^2 + (B^{D}_{n} \delta B^{D}_{n})^2}/a^{D}_{n}, \\
\delta \phi^{D}_{n} = (\cos{\phi^{D}_{n}}/a^{D}_{n})^2 \sqrt{(B^{D}_{n} \delta A^{D}_{n})^2 + (A^{D}_{n} \delta B^{D}_{n})^2},
\end{eqnarray}
where
\begin{eqnarray}
\delta A^{D}_{n} = (\omega/\pi) \sqrt{\sum^{24}_{i=1} {\delta_i}^2 {\cos}^2 \omega t_{i}}, \\
\delta B^{D}_{n} = (\omega/\pi) \sqrt{\sum^{24}_{i=1} {\delta_i}^2 {\sin}^2 \omega t_{i}}.
\end{eqnarray}
We then obtain $a^{R}_{n}$, $\phi^{R}_{n}$ and errors according to eqs. (5) and (6). We also calculate the daily variation, i.e. the 24-hour profile of $R(t)$,  by ``integrating'' $D(t)$ in eq. (2) with respect to $t$ as
\begin{equation}
R^{\ast}(t_{i})=\sum_{k=1}^{i} D(t_{k})-\langle R^{\ast} \rangle,
\end{equation}
where $\langle R^{\ast} \rangle=\sum_{i=1}^{24} \sum_{k=1}^{i} D(t_{k})/24$.
 
Fig.~1 displays $a^{R}_{n}$ and $\phi^{R}_{n}$ for $n=1$ as harmonic vectors in the sidereal (SI) (366.25 days yr$^{-1}$), solar (SO) (365.25 days yr$^{-1}$) and anti-sidereal (ASI) (364.25 days yr$^{-1}$) time frames. The vectors in ASI time frame are nonphysical and are calculated for checking the ``sideband'' effect possibly arising from the seasonal variation in the amplitude of vectors in the SO time frame. It is clear that the diurnal ($n=1$) vectors are most significant in SI and SO time frames, while they are less significant in ASI time frame ensuring the small sideband effect. The amplitude of the semi-diurnal ($n=2$) vector is about one third of the diurnal vector and is significant in SI and SO time frames, while it is insignificant in the ASI time frame. The tri-diurnal ($n=3$) vector is insignificant in all time frames.

The observed diurnal vector in the ASI time frame in Fig. 1 is consistent with the sideband effect expected from the seasonal change of the diurnal vector arising from the second order anisotropy of solar origin \citep{Nagashima83}. The second order anisotropy is symmetric with respect to the interplanetary magnetic field with minimum intensities in directions parallel and anti-parallel to the magnetic field. This anisotropy produces the diurnal variation as well as the semi-diurnal variation in the SO time frame because of the inclination of the earth's spin axis from the magnetic field direction. As the angle between the earth's spin axis and the mean magnetic field is subject to the seasonal variation due to earth's orbital motion around the sun, the amplitude of the diurnal variation also shows the seasonal variation producing spurious diurnal variations in the ASI and SI time frames as sideband effect. In this case, the space harmonic vectors in the ASI and SO time frames are expected to satisfy the following conditions \citep{Nagashima83}.
\begin{equation}
\phi^{R(ASI)}_{1}=23.3 \mbox{  hour},
\end{equation}
\begin{equation}
\phi^{R(SO)}_{2}=3.1 \mbox{  hour},
\end{equation}
\begin{equation}
a^{R(ASI)}_{1}/a^{R(SO)}_{2}=0.38,
\end{equation}
where $\phi^{R(ASI)}_{1}$ ($\phi^{R(SO)}_{2}$) and $a^{R(ASI)}_{1}$ ($a^{R(SO)}_{2}$) denote respectively the phase and amplitude of the diurnal (semi-diurnal) vector in the ASI (SO) time frame in space. The observed average $\phi^{R(ASI)}_{1}$ ($\phi^{R(SO)}_{2}$) in the ASI (SO) time frame over the entire observation period is 2.0$\pm$1.8 hour (2.5$\pm$0.6 hour), while the observed amplitude ratio $a^{R(ASI)}_{1}/a^{R(SO)}_{2}$ is 0.26$\pm$0.15, after the correction for the geomagnetic deflection and the attenuation in the atmosphere. The observed harmonic vectors, therefore, are consistent with the expectation within errors. Based on this, we obtain the spurious harmonic vectors in SI time frame following \citet{Nagashima83}, as
\begin{equation}
a^{R(SI)}_{1}=0.948 a^{R(ASI)}_{1},
\end{equation}
\begin{equation}
\phi^{R(SI)}_{1}=\phi^{R(ASI)}_{1}-4.53 \mbox{  hour},
\end{equation}
\begin{equation}
a^{R(SI)}_{2}=0.098 a^{R(SO)}_{2},
\end{equation}
\begin{equation}
\phi^{R(SI)}_{2}=\phi^{R(SO)}_{2}+4.56 \mbox{  hour}.
\end{equation}
We correct the observed harmonic vectors in the SI time frame by subtracting spurious vectors in eqs.~(19)-(22). Fig.1 displays the corrected and uncorrected SI vectors by full black and full gray circles, respectively. Table 1 presents the corrected $a^{R}_{1}$ and $\phi^{R}_{1}$ in the SI time frame, together with the observed $a^{R}_{1}$ and $\phi^{R}_{1}$ in the SO and ASI time frames for each year between 1985 and 2008. We analyze these results in the following subsections. 

\subsection{Solar cycle variation of the solar diurnal anisotropy}

One remarkable feature of Fig.~1 is that the solar diurnal vector changes periodically every $\sim$11 years. This is evident from Fig.~1 as the summation dial of the diurnal vector in the SO time frame is winding periodically in the second quadrant between 06:00 and 12:00 local time (LT). In our separate paper, we reported that this is due to an ``additional'' anisotropy component superposed on the constant Compton-Getting (CG) anisotropy arising from the earth's orbital motion around the sun \citep{Compton35,Amenomori04,Munakata06}. This ``additional'' anisotropy has a constant phase at $\sim$15:00 local time (LT) and an amplitude changing periodically every $\sim$11 years. The year-to-year  variation of the phase and amplitude of this additional vector, derived by subtracting the expected CG effect from the observed vector, is shown in Fig.~2b. For comparison, we also show in Fig.~2a the yearly mean sunspot number (SSN) and the neutral sheet tilt angle (TA), which are often used as indicators of the solar activity and the solar modulation. The average phase of the additional anisotropy in the 6 years (1991, 1992,1994 and 2002-2004; period I) when the amplitude exceeds 0.035 $\%$ is 15:10$\pm$00:25 LT, while the average amplitude in the same period is 0.043$\pm$0.002 $\%$, with errors deduced from the dispersion of 6 yearly values. On the other hand, the average amplitude in the 6 years (1986-1988, 1997, 1998 and 2007; period II) when the amplitude is smaller than 0.015 $\%$ is 0.008$\pm$0.002 $\%$.  Fig.~3 shows the average solar daily variations observed in period I (a) and period II (b). Plotted in this figure is $R^{\ast}(t_{i})$, obtained by ``integrating'' $D(t)$ in eq.~(15). It is clear that $R^{\ast}(t_{i})$ is fairly consistent with the CG anisotropy in panel (b) with the maximum intensity at $\sim$06:00 LT displayed by a dashed curve, while $R^{\ast}(t_{i})$ in panel (a) deviates from the CG anisotropy showing a maximum intensity at $\sim$12:00 LT due to the additional anisotropy with the maximum at $\sim$15:00 LT superimposed. 

While a clear $\sim$11-year solar cycle variation is seen in the amplitude of the additional anisotropy in Fig.~2b, there is a significant time lag between the temporal variations of the amplitude and the SSN or TA in Fig.~2a. We calculate the 12-month central moving average of the SSN and evaluate the correlation coefficient between the amplitude and the SSN shifting the SSN data to the later period every one month from 0 to +48 months. The time-shifted coefficient gradually increases from +0.25 at 0 month lag, reaching the maximum of +0.74 at +26 month lag and then decreases to +0.26 at +48 month lag. We also calculate the correlation with the TA plotted in Fig.~2a and obtain the maximum correlation coefficient of +0.63 at +30 months. Based on these observational results, we will discuss the possible origin of the additional anisotropy in \S 3.

\subsection{Long-term variation of the sidereal anisotropy}

Fig.~2c shows the year-to-year variation of the amplitude and phase of the sidereal diurnal anisotropy. The average amplitude over the entire period is 0.034$\pm$0.003 $\%$ which is roughly one third of the amplitude reported from AS and deep-underground muon experiments monitoring multi-TeV GCR intensity \citep[e.g., see][]{Nagashima89,Amenomori05,Guillian07,Abdo09,Amenomori09}. This attenuation is probably due to the solar modulation effects, but there is only a weak positive correlation between the amplitude and the SSN and/or TA in this figure. Such a weak positive correlation also has been reported from the long-term observation using an underground muon detector at a shallower depth \citep{Nagashima04}. We obtain the maximum correlation coefficient of +0.57 (+0.50) between the amplitude and the SSN (TA) at the time lag of +14 (+14) months applied to the SSN (TA). We examine the correlation with the solar activity by comparing the average sidereal daily variations $R^{\ast}(t_{i})$ in eq.~(15) in the ``active'' period of 9 years (1988-1992, 1999-2002) when the yearly mean SSN exceeded 90 with that in the ``quiet'' period of 11 years (1985-1987, 1994-1997, 2005-2008) when the SSN was below 30. In order to examine the correlation with the 22-year solar magnetic cycle, we also calculate the average $R^{\ast}(t_{i})$ in the ``positive'' (8 years; 1991-1998) and ``negative'' (12 years; 1985-1988, 2001-2008) periods, separated by the polarity reversal epochs of the solar polar magnetic field indicated in Fig.~2. During the ``positive'' (``negative'') period, the average interplanetary magnetic field (IMF) is directed away from (toward) the sun in the northern hemisphere. Fig.~4a-4d display $R^{\ast}(t_{i})$s in the ``active'', ``quiet'', ``positive'' and ``negative'' periods, respectively. The differences between $R^{\ast}(t_{i})$s in the four periods in Fig.~4 are all insignificant. The amplitude of the daily variation in ``active'' period in Fig.~4a, for instance, appears larger than that in the ``quiet'' period in Fig.~4b, but the intensity difference in each LT remains below three times of the error deduced from the dispersion of yearly mean intensities in each period. The difference between $R^{\ast}(t_{i})$s in ``positive''  and ``negative'' periods in Figs.~4c and 4d remains below two times of the error. It is evident, on the other hand, that the overall feature of $R^{\ast}(t_{i})$ in Fig.~4 with a minimum intensity at $\sim$12:00 LT and a maximum at $\sim$06:00 LT is common for all periods and is quite similar to that observed with AS and deep-underground muon experiments measuring multi-TeV GCRs \citep[e.g., see][]{Nagashima89,Amenomori05,Guillian07,Abdo09,Amenomori09}.

We finally compare the temporal variation of $R(t)$ with that reported by the Milagro experiment. Fig.~5 compares the ``single-band valley depths'' (SBVDs) reported by the Milagro experiment \citep{Abdo09} and those observed by Matsushiro during the identical 7-year period from 2000 to 2007. The SBVD by Matsushiro in this figure is multiplied by three in order to roughly compensate the attenuation of the amplitude in sub-TeV region. We derived the SBVD from Matsushiro data by calculating the minimum intensity of $R(t)$ reconstructed from yearly mean $a^{D}_{n}$ and $\phi^{D}_{n}$ in eq.~(3). Adjusting to the analysis period in \citet{Abdo09}, we calculated the yearly mean SBVD by Matsushiro from 12-month data between June and July. The steady increase in the SBVD reported by the Milagro experiment is not seen in the Matsushiro record. We also calculated SBVD every year in the entire period of the observation by Matsushiro and confirmed that the long-term variation of the SBVD is similar to the variation of the amplitude of the sidereal diurnal anisotropy in Fig.~2c which shows no significant correlation with the solar activity- and magnetic-cycles. We conclude, therefore, that the steady increase of the SBVD reported by Milagro experiment is, most likely, not due to the decreasing solar modulation in the declining phase of the 23rd solar activity cycle.

\section{Discussions}

A clear $\sim$11-year variation was found in the solar diurnal anisotropy. The amplitude of the residual anisotropy, after subtracting the contribution from the Compton-Getting anisotropy due to the earth's orbital motion around the sun, varies in an $\sim$11-year cycle between a maximum amplitude of 0.043$\pm$0.002 $\%$ and a minimum of 0.008$\pm$0.002 $\%$, while its phase remains fairly constant around 15:10$\pm$00:25 LT. A possibility is that the anisotropy is connected with the temporal variation of the IMF which, on large scales, is influenced by the neutral sheet tilt angle (TA). \citet{Erdos80} first suggested that GCRs are expected to experience the energy change when they cross the neutral sheet on their orbital motion through the heliosphere to the earth. They computed the energy change along individual trajectories and deduced the directional anisotropy expected at the earth. Preliminary results of recent numerical model simulations using different tilt angles \citep{Kota07} showed some qualitative similarities to the Matsushiro observations for solar anisotropies: high tilt angles in the maximum period produced larger anisotropies while the phases turned out fairly stable typically in the 3rd quadrant between 12:00 and 18:00 LT. We tentatively note that the $\sim$15:00 phase corresponds to arrival direction of particles that have the best chance to gain energy from the interacting with the corotating magnetic sectors divided by a wavy current sheet. It is most interesting, on the other hand, that the variation of the observed amplitude of the ``additional'' anisotropy, exhibits a significant time lag of +26 months relative to the SSN. This time lag corresponds to the time for the solar wind with an average speed of 400 km/s to propagate through $\sim$180 AU, which is larger than the heliocentric distance of the solar wind termination shock (TS) in the nose direction \citep{Decker05,Stone08}. Even with the slowing down of the subsonic solar wind, this time-lag corresponds to distances deep in the heliosheath, and possibly close to the heliopause \citep{Washimi96}. The cyclic variation induced by the sun propagates to the outer heliosphere with the average solar wind speed. The global distribution of the solar wind plasma and the magnetic field, which governs the TA and the spatial distribution of GCRs, is reorganized when the variation propagates throughout the entire region filled with the solar wind. This implies, therefore, that the intensity of 0.6 TeV GCRs is still a subject to the solar modulation operating over an entire region within the TS and possibly also in the subsonic region beyond the TS. The possible role of heliosheath outside the TS is largely unknown at present and remains to be explored.

We discuss the sidereal anisotropy, which shows no clear correlation with either the solar activity- or magnetic-cycles. Furthermore, there is even a factor of three attenuation of the amplitude relative to air shower observations due to the solar modulation. A possible explanation for this may be found in the region where the attenuation takes place. The solar diurnal anisotropy of 0.6 TeV GCRs at the earth's orbit is mainly produced by the global distribution of the GCR density inside the TS, while the sidereal anisotropy originates from the GCR anisotropy in interstellar space outside the heliosphere and it is attenuated as cosmic rays propagate to the earth through the heliosphere losing their original directional information. If the attenuation occurs predominantly inside the TS, it is reasonable to expect significant solar cycle variation of the anisotropy at the earth's orbit, as we saw in the solar diurnal anisotropy. The situation may change, however, if the attenuation occurs in the region outside the TS. Although our knowledge about the heliosheath between the TS and the heliopause is still limited, the Magneto-Hydrodynamic (MHD) simulations of the interaction between the solar wind and the interstellar plasma suggest the existence of a long heliotail extending over thousand of AU downstream the interstellar plasma flow \citep[e.g., see][]{Baranov93,Washimi96,Zank96,Linde98,Pogorelov04}. Because of both the slower solar wind in the heliosheath and the larger extent of the heliotail, the propagation time of the solar cycle variation through the entire heliotail is expected to take much longer than inside the TS, probably even longer than 11 or 22 years of the solar activity- or magnetic-cycles. In this case, the plasma regions originating from the active and quiet sun are both expected to exist alternatively in the heliotail, as confirmed by a recent MHD simulation \citep{Washimi07}(Dr. H. Washimi, 2009, private communication). If the observed attenuation of the sidereal anisotropy occurs predominantly in the heliotail, therefore, it is possible to expect the attenuation magnitude showing no strong and direct correlation with the solar activity- and/or magnetic-cycles, as we saw in the observation with Matsushiro. This is just one of possible interpretations and we need to confirm it by analyzing the GCR propagation in the model heliosphere including the heliotail predicted by MHD simulations.

Finally we note that the magnetic structure of the heliosheath is largely unknown. The possible role of the heliosheath is particularly interesting in the light of the recent finding of the IBEX spacecraft, mapping the global heliosphere by detecting energetic neutral atoms (ENAs) \citep{McComas09}. The first results of IBEX showed that ENAs are sensitive to, and give information on the structure of the large-scale magnetic field surrounding the heliosphere \citep{Heerik10}. TeV cosmic rays also sense the global structure of the magnetic field within the whole heliosphere as well as the field surrounding the heliosphere, and can provide an additional tool to explore the global structure of these fields.

\acknowledgments

The continuous observation with the Matsushiro underground muon detector has been supported by the Shinshu University and by research grants from the Japanese Ministry of Education, Culture, Sports, Science \& Technology. We thank the Marshall Flight Center, NASA for providing the sunspot number data and the Wilcox Solar Observatory for suppling the neutral sheet tilt angle and the solar polar magnetic field data.

\clearpage
\begin{deluxetable}{lcccccc}
\tabletypesize{\scriptsize}
\tablecaption{Amplitude and phase of the yearly mean diurnal anisotropy observed by Matsushiro in 1985-2008.\label{tbl-1}}
\tablewidth{0pt}
\tablehead{
\multicolumn{1}{c}{} & \multicolumn{2}{c}{SI} & \multicolumn{2}{c}{SO} &\multicolumn{2}{c}{ASI}\\
\multicolumn{1}{c}{\small{year}} & \cline{1-6} \\
\colhead{} & \colhead{$a^{R}_{1}$ ($\%$)} &
\colhead{$\phi^{R}_{1}$ (h)} & \colhead{$a^{R}_{1}$ ($\%$)} &
\colhead{$\phi^{R}_{1}$ (h)} & \colhead{$a^{R}_{1}$ ($\%$)} & \colhead{$\phi^{R}_{1}$ (h)} 
}
\startdata
 1985  &   0.045$\pm$0.013   &   2.66$\pm$2.00   &  0.022$\pm$0.009   &   8.80$\pm$1.58   &  0.009$\pm$0.009   &  23.53$\pm$4.04\\
 1986  &   0.034$\pm$0.012   &   4.57$\pm$1.82   &  0.024$\pm$0.009   &   6.16$\pm$1.40   &  0.011$\pm$0.009   &  23.09$\pm$2.96\\
 1987  &   0.032$\pm$0.013   &  23.53$\pm$0.50   &  0.035$\pm$0.009   &   6.70$\pm$1.01   &  0.003$\pm$0.009   &   7.10$\pm$13.58\\
 1988  &   0.069$\pm$0.012   &   1.38$\pm$1.29   &  0.042$\pm$0.009   &   5.84$\pm$0.80   &  0.013$\pm$0.009   &  16.34$\pm$2.56\\
 1989  &   0.022$\pm$0.012   &  23.66$\pm$0.39   &  0.009$\pm$0.009   &   9.42$\pm$3.89   &  0.019$\pm$0.009   &   6.02$\pm$1.82\\
 1990  &   0.051$\pm$0.013   &   0.95$\pm$0.94   &  0.024$\pm$0.009   &   8.81$\pm$1.50   &  0.012$\pm$0.009   &  10.13$\pm$2.96\\
 1991  &   0.065$\pm$0.013   &   3.98$\pm$1.79   &  0.016$\pm$0.009   &  11.70$\pm$2.29   &  0.025$\pm$0.009   &  22.49$\pm$1.44\\
 1992  &   0.055$\pm$0.012   &   2.96$\pm$2.00   &  0.029$\pm$0.009   &  11.36$\pm$1.21   &  0.014$\pm$0.009   &  15.04$\pm$2.45\\
 1993  &   0.047$\pm$0.013   &   6.01$\pm$1.03   &  0.008$\pm$0.009   &   6.57$\pm$4.54   &  0.022$\pm$0.009   &   2.16$\pm$1.62\\
 1994  &   0.030$\pm$0.013   &  21.77$\pm$1.97   &  0.043$\pm$0.009   &  10.41$\pm$0.84   &  0.028$\pm$0.009   &   7.96$\pm$1.28\\
 1995  &   0.047$\pm$0.013   &   0.13$\pm$0.13   &  0.028$\pm$0.009   &   8.09$\pm$1.25   &  0.002$\pm$0.009   &  12.97$\pm$15.95\\
 1996  &   0.010$\pm$0.013   &  21.46$\pm$3.58   &  0.038$\pm$0.009   &   8.70$\pm$0.93   &  0.013$\pm$0.009   &   8.27$\pm$2.75\\
 1997  &   0.028$\pm$0.013   &   1.95$\pm$1.83   &  0.046$\pm$0.009   &   5.22$\pm$0.75   &  0.012$\pm$0.009   &   7.09$\pm$2.96\\
 1998  &   0.013$\pm$0.013   &  22.64$\pm$1.82   &  0.040$\pm$0.009   &   5.92$\pm$0.87   &  0.018$\pm$0.009   &  13.05$\pm$1.93\\
 1999  &   0.029$\pm$0.013   &   3.09$\pm$2.25   &  0.015$\pm$0.009   &   4.23$\pm$2.34   &  0.022$\pm$0.009   &   2.68$\pm$1.58\\
 2000  &   0.048$\pm$0.013   &   2.06$\pm$1.76   &  0.021$\pm$0.009   &   8.01$\pm$1.70   &  0.016$\pm$0.009   &  18.33$\pm$2.26\\
 2001  &   0.052$\pm$0.013   &   4.51$\pm$1.60   &  0.033$\pm$0.009   &   8.34$\pm$1.06   &  0.030$\pm$0.009   &   0.05$\pm$1.19\\
 2002  &   0.053$\pm$0.013   &   2.15$\pm$1.79   &  0.034$\pm$0.009   &  10.76$\pm$1.03   &  0.021$\pm$0.009   &  20.12$\pm$1.64\\
 2003  &   0.046$\pm$0.013   &   4.35$\pm$1.75   &  0.027$\pm$0.009   &  12.95$\pm$1.30   &  0.033$\pm$0.009   &  22.32$\pm$1.08\\
 2004  &   0.023$\pm$0.013   &   4.28$\pm$2.52   &  0.023$\pm$0.010   &  11.81$\pm$1.61   &  0.007$\pm$0.010   &   3.22$\pm$5.23\\
 2005  &   0.020$\pm$0.014   &  23.98$\pm$0.03   &  0.050$\pm$0.010   &   7.78$\pm$0.79   &  0.013$\pm$0.010   &   7.21$\pm$3.09\\
 2006  &   0.035$\pm$0.013   &   4.39$\pm$1.91   &  0.009$\pm$0.009   &   6.41$\pm$3.79   &  0.008$\pm$0.009   &  22.15$\pm$4.20\\
 2007  &   0.034$\pm$0.013   &   2.48$\pm$2.03   &  0.033$\pm$0.009   &   6.07$\pm$1.07   &  0.009$\pm$0.009   &  23.49$\pm$3.85\\
 2008  &   0.033$\pm$0.013   &   2.48$\pm$2.03   &  0.028$\pm$0.009   &   8.84$\pm$1.30   &  0.003$\pm$0.009   &   1.76$\pm$10.31\\
\enddata
\tablenotetext{a}{Errors are deduced from the muon counts. The amplitude $a^{R}_{1}$ and phase $\phi^{R}_{1}$ in SI time frame are corrected for the ``sideband'' effect (see text).}
\end{deluxetable}

\clearpage
\begin{figure}
\vspace{5mm}
\begin{center}
\includegraphics[scale=0.6,angle=-90]{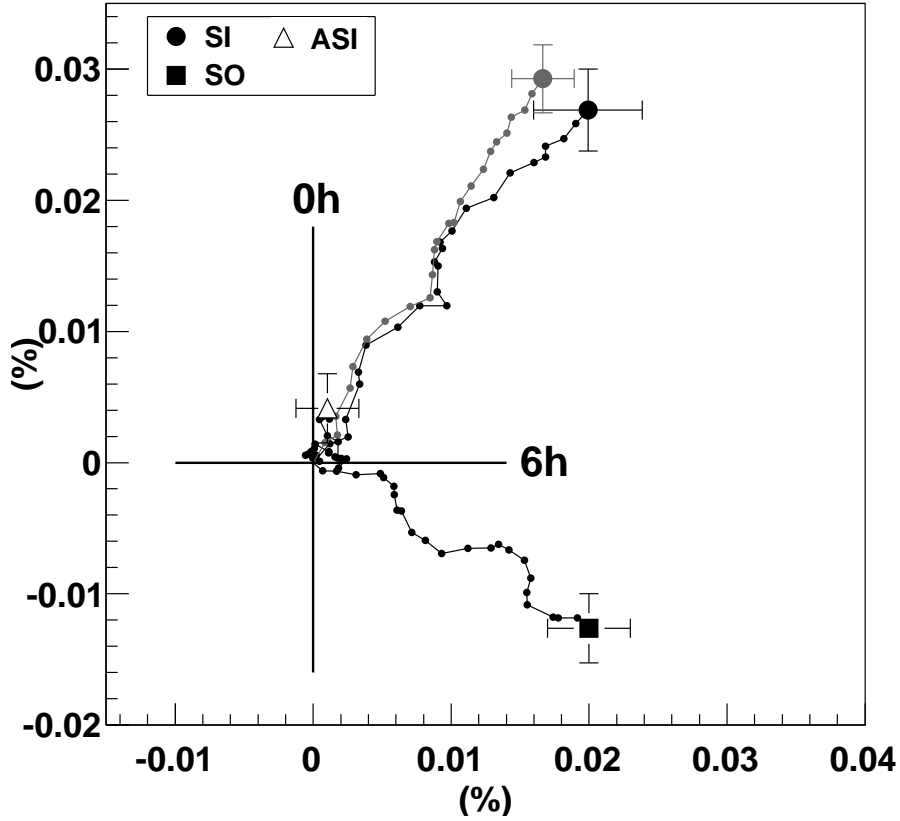}
\caption{
Summation dials of the yearly mean harmonic vector observed by Matsushiro in 1985-2008. Yearly mean diurnal vectors are displayed in the harmonic dial, where the amplitude ($a^{R}_{1}$) of each yearly vector is represented by its length from the origin, while the phase ($\phi^{R}_{1}$) is represented by the angle measured clockwise from the vertical ($+y$) axis, i.e. the local times of $+y$, $+x$, $-y$, and $-x$ axis directions in this figure are 00:00, 06:00, 12:00 and 18:00 in the local time (LT), respectively. To demonstrate the long-term variation, we plot ``vector summation dials'' in this figure where yearly vectors are summed one by one from the first year of 1985. We also divide each yearly vector by 24, the number of years of observation. The final data point in each summation dial, therefore, represents the average yearly vector. The average vectors in the SI, SO and ASI time frames are shown by full circle, full square and open triangle, respectively. The vectors in the SI time frame are corrected for the ``sideband'' effect (see text). The uncorrected average SI vector is also displayed by a gray circle. Error bars of each average vector are deduced from the dispersion of $x$- and $y$-components of 24 yearly mean vectors.
}
\end{center}
\end{figure}

\clearpage
\begin{figure}
\begin{center}
\includegraphics[scale=0.45,angle=0]{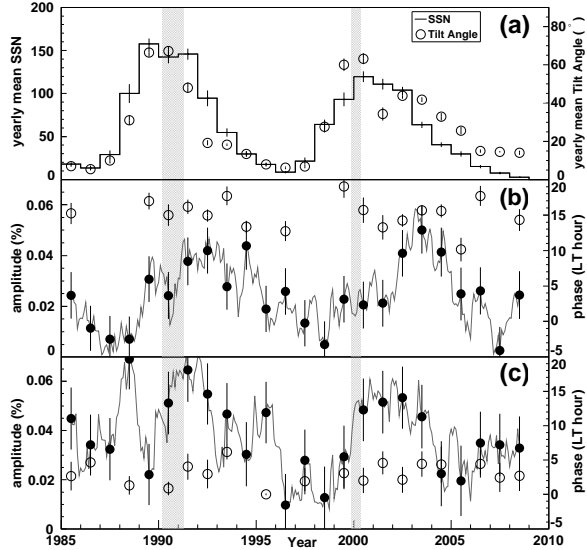}
\caption{
Long-term variation of the diurnal anisotropy observed by Matsushiro in 1985-2008. Yearly mean amplitude (solid circles) and phase (open circles) of the solar diurnal anisotropy (b) and the sidereal diurnal anisotropy (c) are plotted as functions of year. The sidereal diurnal anisotropy in (c) is corrected for the ``sideband'' effect (see text). The yearly averages of the sunspot number (SSN) and the neutral sheet tilt angle (TA) representing the solar activity are also shown in panel (a). Plotted in panel (b) are the amplitude and phase of the ``additional'' anisotropy which is derived by subtracting the harmonic vector expected from the Compton-Getting anisotropy (with an amplitude of 0.035 $\%$ and a phase of 05:56 LT) from the observed anisotropy. The yearly mean phases in years when the amplitude is below 0.02 $\%$ are omitted from the plot in order to exclude large fluctuation due to the insignificant anisotropy. Errors in panels (b) and (c) are statistical errors deduced from the muon counts. The gray solid curves in (b) and (c) also show the 12-month central moving average of the amplitude. The yearly mean SSN and TA are calculated respectively from the monthly SSN value (http://solarscience.msfc.nasa.gov) and the Carrington rotation average of the TA computed using a model based on the radial boundary condition at the photosphere (http://wso.stanford.edu/Tilts.html). Errors in panel (a) are deduced from dispersions of the monthly mean SSN and the Carrington rotation average of TA used for calculating yearly mean values. The periods of the solar polar magnetic field reversals deduced from the solar polar magnetic field data (http://wso.stanford.edu/Polar.html) are also indicated by the vertical shaded epochs.
}
\end{center}
\end{figure}

\clearpage
\begin{figure}
\begin{center}
\includegraphics[scale=0.6,angle=0]{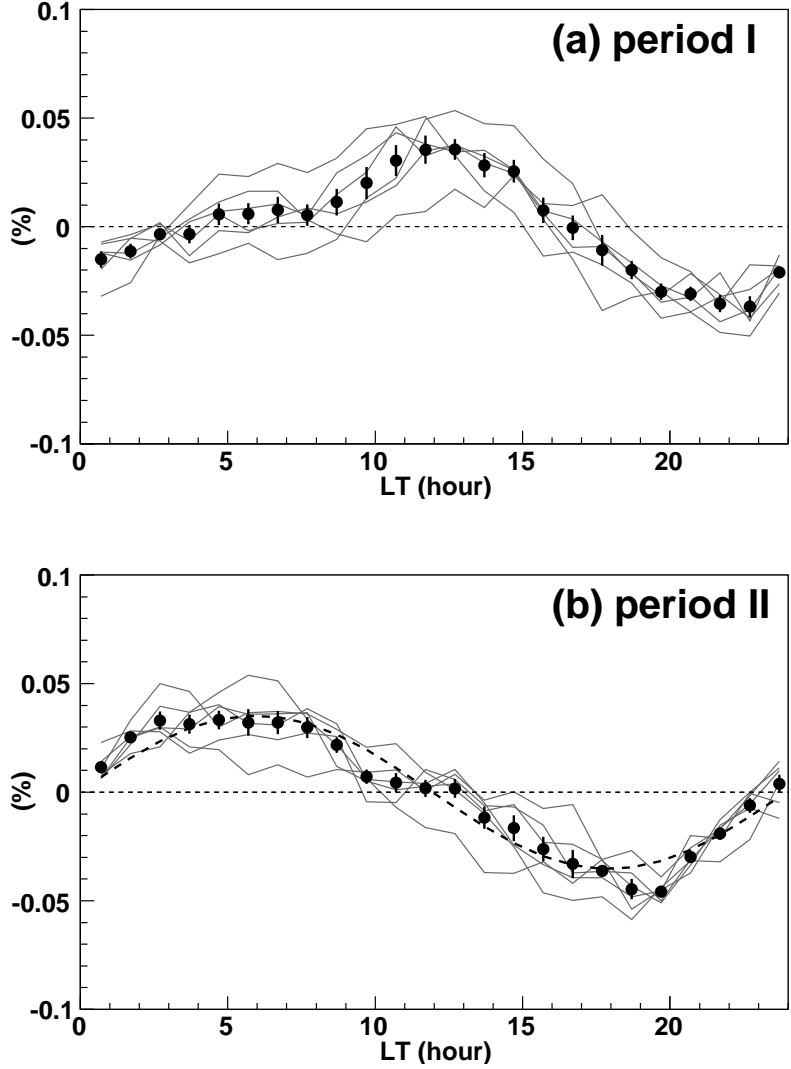}
\caption{
Average solar daily variations $R^{\ast}(t_{i})$ in two periods selected according to the amplitude of the additional anisotropy. Panel (a) shows by solid circles the average $R^{\ast}(t_{i})$ in 6 years of 1991, 1992, 1994 and 2002-2004 (period I) when the amplitude of the ``additional'' anisotropy in Fig.~2b exceeds 0.035 $\%$, while panel (b) displays $R^{\ast}(t_{i})$ in 6 years of 1986-1988, 1997, 1998 and 2007 (period II) when the amplitude is below 0.015 $\%$. Dashed curve in panel (b) represents the diurnal variation expected from the CG anisotropy arising from the earth's orbital motion around the sun. In panels (a) and (b), $R^{\ast}(t_{i})$s are shown as functions of the local solar time $t_{i}$ on the horizontal axis. Error is deduced from the dispersion of the yearly mean of $R^{\ast}(t_{i})$ used for the calculation of the average and displayed by gray curves.
}
\end{center}
\end{figure}

\clearpage
\begin{figure}
\begin{center}
\includegraphics[scale=0.6,angle=-90]{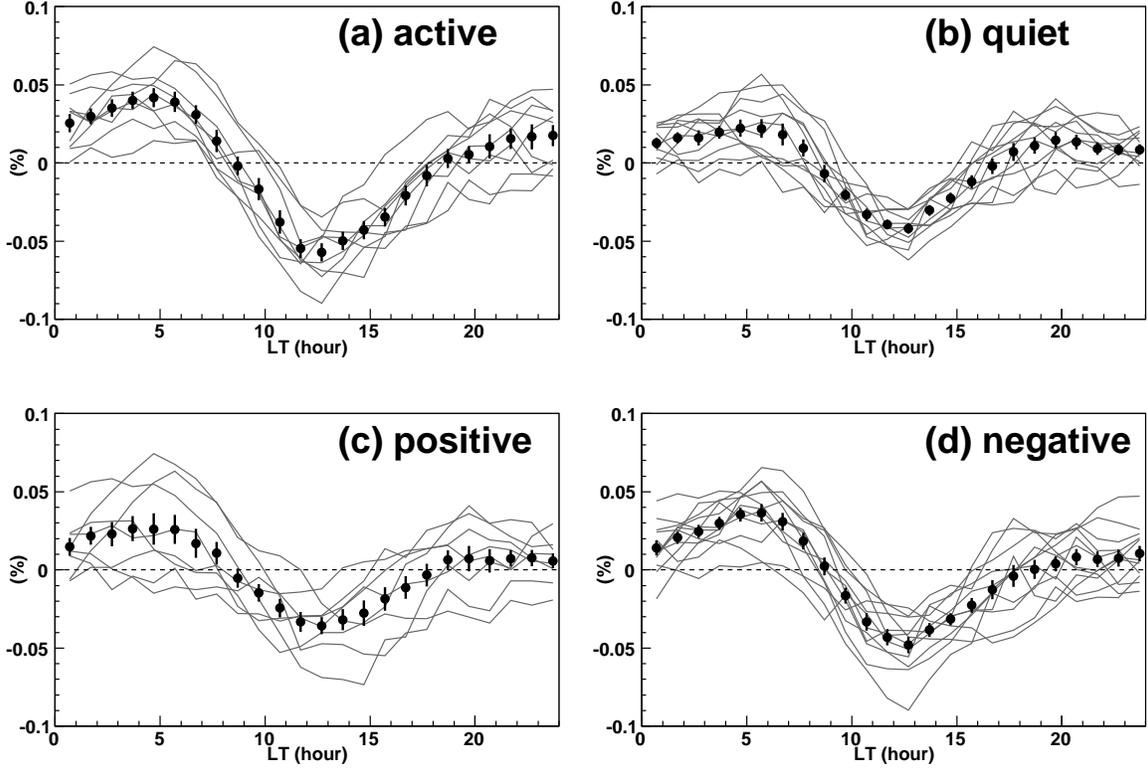}
\caption{
Average sidereal daily variations $R^{\ast}(t_{i})$ in four periods selected according to the solar activity- and magnetic-cycles. All daily variations in this figure are corrected for the ``sideband'' effect (see text). Panels (a) and (b) display by solid circles $R^{\ast}(t_{i})$s in ``active'' (1988-1992, 1999-2002) and ``quiet'' (1985-1987, 1994-1997, 2005-2008) periods of the solar activity, respectively, while panels (c) and (d) show respectively $R^{\ast}(t_{i})$s in ``positive'' (1991-1998) and ``negative'' (1985-1988, 2001-2008) periods of the solar polar magnetic field. Error is deduced from the dispersion of the yearly mean of $R^{\ast}(t_{i})$ used for the calculation of each average and displayed by gray curves.
}
\end{center}
\end{figure}

\clearpage
\begin{figure}
\begin{center}
\includegraphics[scale=0.6,angle=-90]{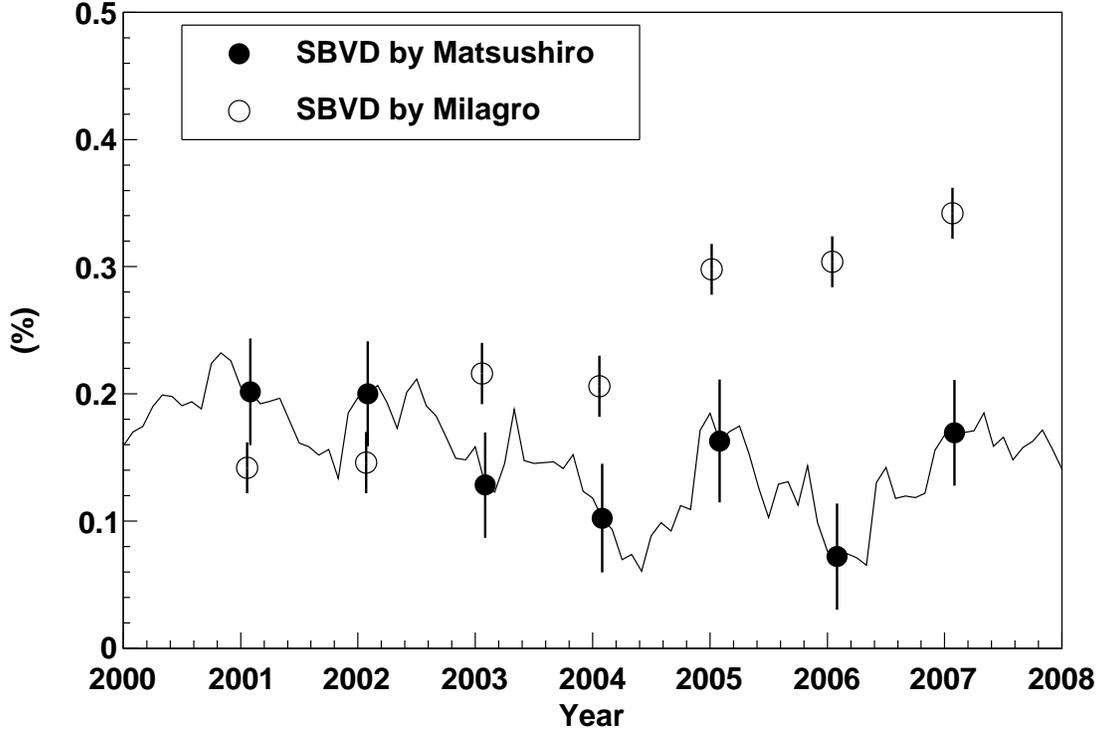}
\caption{
The Single Band Valley Depths (SBVDs) observed by Matsushiro and Milagro experiment in a 7-year period between 2000-2007. The SBVDs by Matsushiro and Milagro experiment are plotted by full and open circles, respectively, as functions of year on the horizontal axis. The SBVD by Matsushiro in this figure is corrected for the ``sideband'' effect and multiplied by three in order to compensate for the attenuation of the amplitude in sub-TeV region. The yearly mean SBVD by Matsushiro is calculated from 12-month data between June and July adjusting to the analysis period by Milagro experiment (Abdo et al., 2009). The gray solid curve also displays the 12-month central moving average of the SBVD by Matsushiro. The error of Matsushiro's SBVD which is corrected for the ``sideband'' effect is only statistical, while the error of Milagro's SBVD includes both the statistical and systematic errors evaluated from the amplitude of the observed ASI diurnal variation \citep[see][]{Abdo09}.  
}
\end{center}
\end{figure}


\begin{thebibliography}{}
\bibitem[Abdo et al. (2009)]{Abdo09}Abdo, A. A., et al. 2009, \apj, 698, 2121-2130
\bibitem[Amenomori et al. (2004)]{Amenomori04}Amenomori, M., et al. 2004, \prl, 93, 061101-1-061101-4
\bibitem[Amenomori et al. (2005)]{Amenomori05}Amenomori, M., et al. 2005, \apjl, 626, L29-L32
\bibitem[Amenomori et al. (2006)]{Amenomori06}Amenomori, M., et al. 2006, Science, 314, 439-443
\bibitem[Amenomori et al. (2009)]{Amenomori09}Amenomori, M., et al. 2009, Advances in Geosciences, in press (arXiv0811.0422)
\bibitem[Amenomori et al. (2009a)]{Amenomori09a}Amenomori, M., et al. 2009a, Proc. 31st Int. Cosmic Ray Conf., paper ID0827, 1-4
\bibitem[Baranov \& Malama (1993)]{Baranov93}Baranov, V. B., \& Malama, Y. G. 1993, \jgr, 98, 15157-15163
\bibitem[Compton \& Getting (1935)]{Compton35}Compton, A. H., \& Getting, I. A. 1935, Phys. Rev., 47, 817-10433
\bibitem[Decker et al. (2005)]{Decker05}Decker, R. B., et al. 2005, Science, 309, 2020-2024
\bibitem[Erd\"{o}s \& K\'{o}ta (1980)]{Erdos80}Erd\"{o}s, G., \& K\'{o}ta, J. 1980, \apss, 67, 45-59
\bibitem[Guillian et al. (2007)]{Guillian07}Guillian, G., et al. 2007, \prd, 75, 062003-1-062003-17
\bibitem[Heerikhuisen et al. (2010)]{Heerik10}Heerikhuisen, J., et al. 2010, \apjl, 708, L126-L130
\bibitem[K\'{o}ta et al. (2007)]{Kota07}K\'{o}ta, J., et al., 2007, Proc. 30th Int. Cosmic Ray Conf., 1 589-592
\bibitem[Lin et al. (1995)]{Lin95}Lin, Z., et al. 1995, \jgr, 100, 23543-23549
\bibitem[Linde et al. (1998)]{Linde98}Linde, T. J., et al., 1998, \jgr, 103, 1889-1904
\bibitem[McComas et al. (2009)]{McComas09}McComas, D. J., et al. 2009, Science Express Report, 10.1126/science.1180927
\bibitem[Munakata et al. (1997)]{Munakata97}Munakata, K., et al. 1997, \prd, 56, 23-26
\bibitem[Munakata et al. (2006)]{Munakata06}Munakata, K., et al. 2006, Advances in Geosciences, 2, 125-134 (reprint is available at http://cosray.shinshu-u.ac.jp/crest/)
\bibitem[Murakami et al. (1979)]{Murakami79}Murakami, K., et al. 1979, IL Nuovo Cimento, 2C, 635-651
\bibitem[Nagashima et al. (1983)]{Nagashima83}Nagashima, K., et al. 1983, IL Nuovo Cimento, 6C, 550-565
\bibitem[Nagashima et al. (1989)]{Nagashima89}Nagashima, K., et al. 1989, IL Nuovo Cimento, 12C, 695-749
\bibitem[Nagashima et al. (1998)]{Nagashima98}Nagashima, K., et al. 1998, \jgr, 103, 17429-17440
\bibitem[Nagashima et al. (2004)]{Nagashima04}Nagashima, K., et al. 2004, Earth Planets Space, 56, 479-483
\bibitem[Pogorelov et al. (2004)]{Pogorelov04}Pogorelov, N., et al. 2004, \apj, 614, 1007-1021
\bibitem[Stone et al. (2008)]{Stone08}Stone, E. C., et al. 2008, \nat, 453, 71-74
\bibitem[Washimi \& Tanaka (1996)]{Washimi96}Washimi, H., \& Tanaka, T. 1996, \ssr, 78, 85-94
\bibitem[Washimi et al. (2007)]{Washimi07}Washimi, H., et al. 2007, \apjl, 670, L139-L142
\bibitem[Zank et al. (1996)]{Zank96}Zank, G. P., et al., 1996, \jgr, 101, 21639-21655
\end{thebibliography}
\end{document}